\documentclass[12pt,preprint]{aastex}
\usepackage{epsfig}

\newcommand{\simgt}{\lower.5ex\hbox{$\;\buildrel>\over\sim\;$}}
\newcommand{\simlt}{\lower.5ex\hbox{$\;\buildrel<\over\sim\;$}}
\newcommand{\zsun}{\ensuremath{Z_\odot}}
\newcommand{\msun}{\ensuremath{~M_\odot}}

\newcommand{\hi}{~H{\sc I}}
\newcommand{\hii}{~H{\sc II}}
\newcommand{\av}{\ensuremath{~A_V}}
\newcommand{\nuo}{{\ensuremath{\nu_0}}}
\newcommand{\alphant}{\ensuremath{\alpha_{\mathrm{nt}}}}
\newcommand{\tauff}{\ensuremath{\tau_{\mathrm{ff}}}}
\newcommand{\nion}{N$_{\rm ion}$}

\newcommand{\lnt}{$L_{\rm NT}$}
\newcommand{\lsyn}{$L_{\rm syn}$}
\newcommand{\lt}{$L_{\rm T}$}
\newcommand{\snr}{$\nu_{\rm SN}$}
\newcommand{\bra}{${\rm Br}\alpha$}
\newcommand{\brg}{${\rm Br}\gamma$}
\newcommand{\ha}{${\rm H}\alpha$}
\newcommand{\hb}{${\rm H}\beta$}
\newcommand{\lya}{${\rm Ly}\alpha$}

\newcommand{\sbs}{SBS\,0335$-$052}
\newcommand{\izw}{I\,Zw\,18}

\shorttitle{Radio Continuum of \sbs}
\shortauthors{Hunt et al.}

\received{2003 December 3}
\begin{document}

\title{The Radio Continuum of the Metal-Deficient Blue Compact Dwarf Galaxy \sbs}

\author{
	Leslie~K.~Hunt\altaffilmark{1},
        Kristy~K.~Dyer\altaffilmark{2,3},
	Trinh~X.~Thuan\altaffilmark{4}, and
	James~S.~Ulvestad\altaffilmark{2}
}
\altaffiltext{1}{INAF-Istituto di Radioastronomia-Sez.\ Firenze, 
L.go Fermi 5, I-50125 Firenze, Italy; hunt@arcetri.astro.it.}
\altaffiltext{2}{National Radio Astronomy Observatory, P.O. Box O,
Socorro, NM 87801, USA: kdyer@aoc.nrao.edu, julvesta@cv3.cv.nrao.edu}
\altaffiltext{3}{National Science Foundation Astronomy and Astrophysics
Postdoctoral Fellow}
\altaffiltext{4}{Astronomy Department, University of Virginia, Charlottesville, VA 22903, USA: txt@virginia.edu}

\begin{abstract}
We present new Very Large Array observations at five frequencies,
from 1.4 to 22\,GHz, of the extremely
low-metallicity blue compact dwarf \sbs.
The radio spectrum shows considerable absorption at 1.49\,GHz,
and a composite thermal$+$non-thermal slope.
After fitting the data with a variety of models,
we find the best-fitting geometry to be one with free-free
absorption homogeneously intermixed with the emission
of both thermal and non-thermal components.
The best-fitting model gives an an emission measure
$EM\sim8\times10^7$\,pc\,cm$^{-6}$
and a diameter of the radio-emitting region $D\,\approx$\,17\,pc. 
The inferred density is $n_e\,\sim\,2000$\,cm$^{-3}$.
The thermal emission comes from an ensemble of $\sim\,9000$\,O7 stars,
with a massive star-formation rate ($\geq5\,M_\odot$) of 0.13-0.15\,yr$^{-1}$,
and a supernova rate of 0.006\,yr$^{-1}$.
We find evidence for ionized gas emission from stellar winds, since the
observed \bra\ line flux significantly exceeds that inferred from the 
thermal radio emission.
The non-thermal fraction at 5\,GHz is $\sim$0.7, corresponding to a
non-thermal luminosity of $\sim2\times10^{20}$\,W\,Hz$^{-1}$.
We derive an equipartition magnetic field of $\sim$\,0.6-1\,mG,
and a pressure of $\sim\,3\times10^{-8}-1\times10^{-7}$\,dyne\,cm$^{-2}$.
Because of the young age and compact size of the starburst,
it is difficult to interpret the non-thermal radio emission as resulting
from diffusion of 
supernova (SN) accelerated electrons over $10^7-10^8$\,yr timescales.
Rather, we attribute the non-thermal radio emission 
to an
ensemble of compact SN remnants expanding in a dense interstellar medium.
If the radio properties of \sbs\ are representative of star formation
in extremely low-metallicity environments, derivations of the star formation
rate from the radio continuum in high redshift primordial galaxies 
need to be reconsidered.
Moreover, photometric redshifts inferred from ``standard'' spectral energy
distributions could be incorrect.
\end{abstract}


\section{\label{intro}Introduction}

When and how the first episodes of star formation took place remains
one of the main questions of modern cosmology.
Much effort has been devoted to measuring the cosmic star formation rate
(SFR) of the universe as a function of redshift or look-back time
initially at ultraviolet wavelengths (e.g., \citealt{madau}),
and more recently in the millimeter and radio regimes
(e.g., \citealt{blain,haarsma}). 
Unlike ultraviolet or optical wavelengths,
radio emission at \simgt\ 1~GHz is unaffected by dust,
which eliminates the need for uncertain extinction corrections.
Nevertheless, the conversion of radio luminosity to SFR depends on
canonical scaling relations (e.g., \citealt{condon}) which may not 
apply to star formation at high redshift.

Primordial star formation occurs in chemically unenriched environments.
Besides the high-redshift \lya\ absorption systems which are very difficult
to study,
the only examples we have of such environments are
nearby metal-poor galaxies actively forming stars, such as the
class of galaxies known as Blue Compact Dwarfs (BCDs). 
Unlike solar-metallicity star-forming regions,
the radio emission in these objects tends to be predominantly thermal rather
than non-thermal \citep{klein84,klein91}.
Low-metallicity BCDs also tend to be of low mass.
Hence, they may represent the primordial ``building blocks'' --
or ``sub-galaxies'' \citep{rees} --
in hierarchical scenarios of galaxy formation.
As such, the study of their radio properties can be useful to better
understand primordial galaxies.

One of the best candidates for the primordial building-block status is \sbs.
At $Z\,\sim\,$1/41 \zsun\ \citep{melnick,izotov97}, it is the lowest-metallicity
star-forming galaxy known after \izw\ ($Z\,\sim\,$1/50 \zsun).
Its mass in stars is $\simlt\ 10^7$\msun\ \citep{papaderos,vanzi}, and it is
embedded in an \hi\ cloud of $\sim\ 2\times 10^9$\msun\ \citep{pustilnik}.
To explore the radio properties of star formation in a chemically
unenriched interstellar medium,
and to better understand the nature of the starburst in \sbs,
we have obtained Very Large Array (VLA) D-array observations at five 
continuum frequencies.
In this paper, we first describe in some detail the extremely
metal-poor BCD \sbs.
We then present the new radio observations and their reduction
and analysis in $\S$\ref{observations}, together with
a description of the optical-radio alignment. 
Models of the continuum emission and the results of the spectral fitting 
are discussed in $\S$\ref{models} and $\S$\ref{results}, respectively.
In $\S$\ref{sfr} we examine the properties of the starburst in \sbs, as inferred
from our observations, and assess the nature of the non-thermal
radio source.
Finally, we discuss the possible implications of our results for primordial 
star formation.

\section{\sbs}

\begin{figure}
\centerline{
\epsfig{file=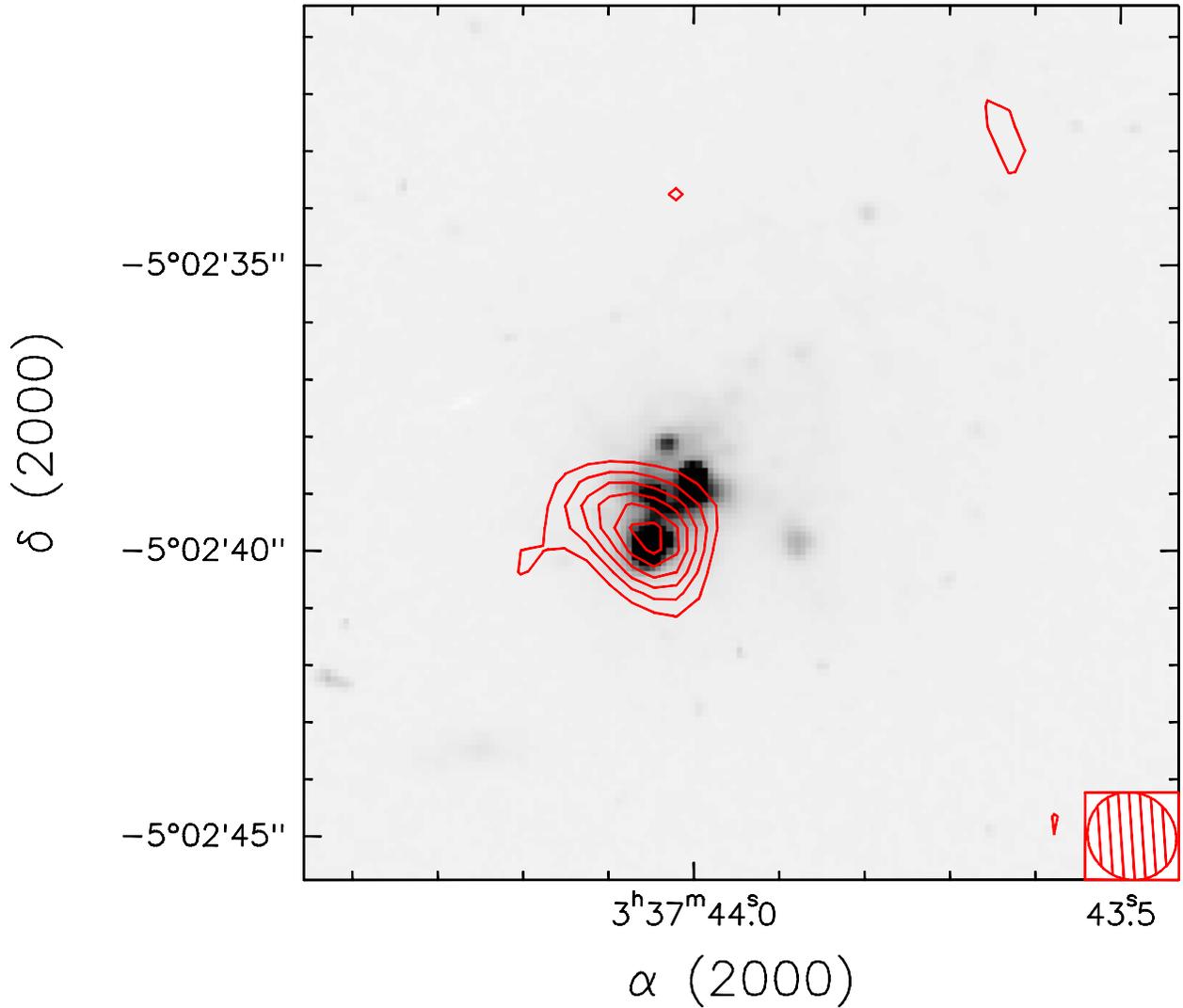, angle=270, width=\linewidth}
}
\caption{20\,cm map overlayed on the HST/WFPC2 (F791W) image.
Radio contours run from 0.12\,mJy/beam (2$\sigma$) to 0.42\,mJy/beam
in units of $\sigma$ (0.06\,mJy/beam).
The beam, 1.6\arcsec$\times$1.5\arcsec, is shown in the lower right
corner.
The radio source is clearly coincident with the two brightest
SSCs in the southeast portion of \sbs.
\label{fig:overlay}}
\end{figure}

At $Z/\zsun\,\sim$\,1/41,
\sbs\ is the lowest-metallicity galaxy in the Second Byurakan Survey 
\citep{sbs-survey}.
\sbs\ comprises six main star clusters surrounded by
a very blue ($V-I\,=\,-0.3-0.2$) low-surface brightness 
diffuse component which, because of its filamentary structure and
color, is almost certainly gas \citep{thuan97,papaderos}.
It is more distant (redshift distance is 54.3\,Mpc), and 
fainter ($V$\,=\,16.65) than \izw, but more 
luminous (M$_B$\,=\,-16.7, \citealt{thuan97}).
\sbs\ has a companion of similar metal abundance $\sim\,$20\,kpc to the west,
and both are embedded within a common \hi\ cloud \citep{pustilnik}.
An HST (F791W) image of \sbs\ is shown in Fig. \ref{fig:overlay}.

The star clusters in \sbs\ are sufficiently compact 
(diameter upper limit $\leq\,$50\,pc) and luminous 
($-14.1\,\leq\,V\,\leq\,-11.9$) to be defined as Super Star Clusters (SSCs) \citep{thuan97}.
Massive star density in the SSCs is $\simgt\,$2.5\,pc$^{-2}$, with a high 
surface brightness 
$\simlt$\,16\,mag\,arcsec$^{-2}$\ in $K$ for the two brightest SSCs \citep{vanzi}. 
Optical spectra of \sbs\ show evidence for Wolf-Rayet stars
\citep{guseva}, and a high electron temperature 
(20000--22000\,K) and density
(500--600\,cm$^{-3}$, \citealt{izotov99}).

Extinction as measured from the optical hydrogen recombination lines in the 
brightest SSC pair is \av\,=\,0.55 mag \citep{izotov97},
but somewhat higher if measured with \brg: \av\,=\,0.73 mag \citep{vanzi}.
Extinction is higher still if measured with \bra\ (\hb/\bra: 
\av\,=\,1.5 mag; \brg/\bra: \av\,=\,12.1 mag; \citealt{hunt}).
Such high extinction appears to be caused by warm dust whose emission 
has been measured in the mid-infrared with the Infrared Space Observatory \citep{thuan99}
and from the ground \citep{dale}. 
The warm dust emitting in the mid-infrared
is associated with the two brightest SSCs, 
and resides in a region no more than 80-100\,pc in diameter
\citep{hunt,dale}; the dust mass is $\simlt\,10^5$\,\msun\ \citep{plante}.
The SFR as derived from the integrated \ha\ emission is 0.4\,\msun/yr
\citep{thuan97}, while from the \bra\ luminosity in
the spectroscopic slit alone is 1.7\msun/yr \citep{hunt}.
Those authors concluded that optical observations do not detect roughly
3/4 of the star formation in the brightest SSCs, and even 2$\mu$m
observations miss half of it.

\section{VLA Observations and Data Reduction\label{observations}}

Exploratory observations of \sbs\/ were taken on 9 November 2001 in D
array at 8.4 GHz resulting in a detection of \sbs\/ in 3.5 hours of
observation. A second round of observations at 1.4, 4.8, 8.5, 15 and
22 GHz were taken on 9
January 2002 during D$\rightarrow$A move. While no source was detected
at 1.4 GHz in 10 minutes, \citet{dale} report a detection of \sbs\/ at
3$\sigma$ in a 1994 observation. We requested and received discretionary
observing time on 12 September 2003 during the A$\rightarrow$BnA move to confirm
this detection and detected \sbs\/ at the 7$\sigma$ level. 

There is no conclusive evidence that \sbs\/ is resolved in this
1.6\arcsec$\times$1.5\arcsec\
resolution image, and we therefore assume it is unresolved in the images
made with a larger beam at other frequencies.
This assumption is corroborated by the excellent agreement of our A$\rightarrow$BnA
flux (0.47\,$\pm$\,0.06 mJy/beam) and that obtained by \citet{dale} at 1.49\,GHz 
(0.40\,mJy/beam) with a 6\arcsec\ beam in the B configuration.
Table \ref{tab:VLA} gives the fluxes, their uncertainties, the beam
sizes and position angle, and the integration time of each observation.
The uncertainty on the flux given in Table \ref{tab:VLA} includes the rms noise and 5\% flux
calibration uncertainty. 

\clearpage
\begin{deluxetable}{rcccc}
\footnotesize
\tablecaption{VLA Observations of \sbs\  \label{tab:VLA}}
\tablewidth{0pt}
\tablehead{
\colhead{Frequency [GHz]}&\colhead{Flux [mJy beam$^{-1}$]}
&\colhead{Beam}&\colhead{Time on source}}
\startdata
1.46 	&0.4571 $\pm$ 0.0610	&1.6\arcsec $\times$1.5\arcsec, 49\arcdeg	&45m\\
4.86  	&0.7720 $\pm$ 0.0855	&29\arcsec $\times$16\arcsec, 54\arcdeg	&12m\\
8.46  	&0.6508 $\pm$ 0.0386	&12\arcsec $\times$9\arcsec, -7\arcdeg		&3.5h\\
14.94 	&0.3960 $\pm$ 0.0704	&12\arcsec $\times$5\arcsec, 54\arcdeg		&1.8h\\
22.46 	&0.4817	$\pm$ 0.0789	&6\arcsec $\times$4\arcsec, 52\arcdeg		&1.3h\\
\enddata
\end{deluxetable}    
\clearpage
\subsection{Optical-Radio Alignment }

The radio source in our highest-resolution (1.5\arcsec$\times$1.6\arcsec)
image at 1.46\,GHz is shown in Figure \ref{fig:overlay}.
The radio contours in units of $\sigma$ are overlayed on the HST/WFPC2
image \citep{thuan97} in the F791W filter which we obtained from the 
HST archive and re-reduced.
We performed our own
astrometrical calibration using stars from the U.S. Naval Observatory
Astrometric Catalog A2.0 
(USNOA2.0\footnote{\url{http://tdc-www.harvard.edu/software/catalogs/ua2.html}}).
In the 1600$\times$1600 pixel WFPC2 image, there are five stars in
the USNOA2.0. 
The astrometric solution for the image, based on these stars, was
calculated with the {\it imwcs} routine in the WCSTools package 
(available from \url{http://tdc-www.harvard.edu/software/wcstools/}).
The solution has an rms uncertainty of 0.54\arcsec, or roughly 5 WFPC2
(mosaic) pixels.

The correct HST/WFPC2 astrometry results in a virtually perfect
alignment of the compact 20\,cm  radio source with the brightest
SSCs ($1+2$) toward the southeast edge of \sbs.
To the level of $\sim\,60\,\mu$Jy/beam,
there is no significant radio emission coming from the other SSCs or
the low-surface brightness envelope surrounding the star clusters.

\section{Models for the Radio Spectrum\label{models}}

The radio continuum spectrum of \sbs\ is nearly flat with a small
negative slope for $\nu>5$\,GHz, and a drop-off at $\nu=1.4$\,GHz
(Figure \ref{fig:fit});
the low-frequency drop-off is a signature of free-free absorption.
The apparent spectral index of $-0.4$ of the four
points with $\nu\,>\,1.4\,$GHz
is similar to that obtained for a large sample of BCDs \citep{klein84} ($\alpha$
from 1.4$-$5\,GHz\,=$-0.33$), implying that absorption in
most other BCDs must be less than in \sbs.
The apparent spectral index is also steeper than optically thin thermal emission, 
implying that there must be some non-thermal contribution to the
observed flux. 

Because of the low-frequency drop,
we did not attempt to fit the radio spectrum of \sbs\ with
the simplest model of combined
thermal$+$non-thermal emission with no absorption
(c.f., \citealt{klein91}).
The models we investigated include thermal and non-thermal
components, and the effects of free-free absorption from
ionized gas with several geometries. 
If \tauff\ is the free-free absorption optical depth
\footnote{This approximation for \tauff\ is accurate to $\simlt$20\%.}:
\begin{equation}
\tauff\,\simeq\,0.08235\ \left(\frac{T}{K}\right)^{-1.35}\ 
\left(\frac{EM}{\rm pc\ cm^{-6}}\right)
\left(\frac{\nu}{\rm GHz}\right)^{-2.1}
\end{equation}
where $\nu$ is the frequency, $T$ is the ionized gas temperature, 
and $EM$ is the emission measure, 
the optically thin thermal emission $f^{\rm th}$ can be written as:
\begin{equation}
f^{\rm th}_\nu\ =\ \phi\ (5.95\times10^{-5})\
\left(\frac{T}{K}\right)^{-0.35} 
\left(\frac{\nu}{\rm GHz}\right)^{-0.1}
\left(\frac{EM}{\rm pc\ cm^{-6}}\right)
\left(\frac{\theta}{\arcsec}\right)^2 \ {\rm mJy}
\end{equation}
$\phi$ is a geometrical factor that is equal to $\pi/6$ for a spherical
region of constant density and diameter $\theta$, and
$\pi/4$ for a cylindrical region of diameter and length $\theta$
(e.g., \citealt{mezger}).
$\phi$ also includes the filling factor, which we have assumed equal to unity.
The non-thermal emission can be written as:
\begin{equation}
f^{\rm nt}_\nu\ = \ f^{\rm nt}_{\nuo}\left(\frac{\nu}{\nuo}\right)^{\alphant} \ {\rm mJy}
\end{equation}
where $f^{\rm nt}_{\nuo}$ is the non-thermal (unabsorbed) flux at \nuo.
Model 1 includes absorption in the form of a foreground screen
of ionized gas that obscures both thermal and non-thermal components:
\begin{equation}
f^{tot}_\nu\ =\ \exp(-\tauff) ( f^{\rm nt}_\nu\ +\ f^{\rm th}_\nu ) \quad\quad {\rm mJy}
\end{equation} 
Model 2 assumes that the absorbing medium is intermixed with both
thermal and non-thermal emission:
\begin{equation}
f^{tot}_\nu\ =\ \left[\frac{1 - \exp(-\tauff)}{\tauff}\right]\ 
(f^{\rm nt}_\nu\ +\ f^{\rm th}_\nu ) \quad\quad {\rm mJy}
\end{equation} 
Model 3 includes a screen absorption term only for the non-thermal
component, while the absorbing medium is assumed to be homogeneously 
intermixed with the thermal emission: 
\begin{equation}
f^{tot}_\nu\ =\ \exp(-\tauff)\ f^{\rm nt}_\nu\ + 
\left[\frac{1 - \exp(-\tauff)}{\tauff}\right]\ f^{\rm th}_\nu  \quad\quad {\rm mJy}
\end{equation} 

These models represent slightly different geometries. 
In models 1 and 2, thermal and non-thermal regions are assumed to be cospatial. 
In model 3, the absorbing medium and thermal emitting region must lie 
between the non-thermal emission and the telescope, or ``outside'' of it.
Models in which the non-thermal emission was not absorbed, for example
when the non-thermal source is more extended than the thermal emitter/absorber,
give significantly worse fits than the three models considered here.

We fit the VLA spectrum to these models using a $\chi^2$ minimization
technique to estimate $f^{\rm nt}_{\nuo}$, $\theta$, and $EM$ (through \tauff)
based on the {\it amoeba} algorithm \citep{press}.
Errors for the fitted parameters were obtained according to the precepts of
\citet{lampton} for three fitted parameters and a confidence level of 95\%.
\nuo\ was chosen to be 5\,GHz.
\alphant\ was fixed {\it a priori} to $-0.8$, although values ranging
from $-0.5$ to $-1.1$ did not affect the relative quality of the fits
(but see $\S$\ref{results}).
The electron temperature $T_e$ was taken to be 20000\,K \citep{izotov99}. 
We have tested our fitting procedure against  \citet{deeg}
for II\,Zw\,40, and obtain results consistent with theirs.

\subsection{Additional Physical Processes}

In addition to the models already described, we also investigated
other physical mechanisms that may help shape the radio
spectrum of \sbs.


Since the ultraviolet (UV) radiation field in \sbs\ is roughly 10000
times the local value \citep{dale}, inverse Compton losses may be
important. 
Following \citet{deeg}, we included this effect in the expression for the
non-thermal spectrum.
These fits gave a larger $\chi^2$, probably
because of the insensitivity of the spectrum to the non-thermal emission,
given the dominant thermal emission at high frequencies in \sbs.


We also investigated the possible effect of a thermal plasma
in the synchrotron emission region.
This effect, usually called the Razin-Tsytovich effect (e.g., \citealt{simon}),
does not appear to influence the shape of \sbs's radio spectrum.
For \sbs, the Razin-Tsytovich cutoff frequency appears to be less than the
thermal cutoff, since we were unable to obtain good fits by including
this effect.
This would be the case if the magnetic field $B\,>\,4\times10^{-5}/\sqrt{L}$
where $L$ is the physical dimension of the emitting region in pc
(see \citealt{verschuur}).
For $L\,=\,30\,$pc, $B\,>\,7\,\mu$G.
The equipartition magnetic field inferred from our observations 
(see $\S$\ref{results}) is more than 100 times this value.

\section{Spectral Fitting Results\label{results}}

\begin{figure}
\centerline{
\plottwo{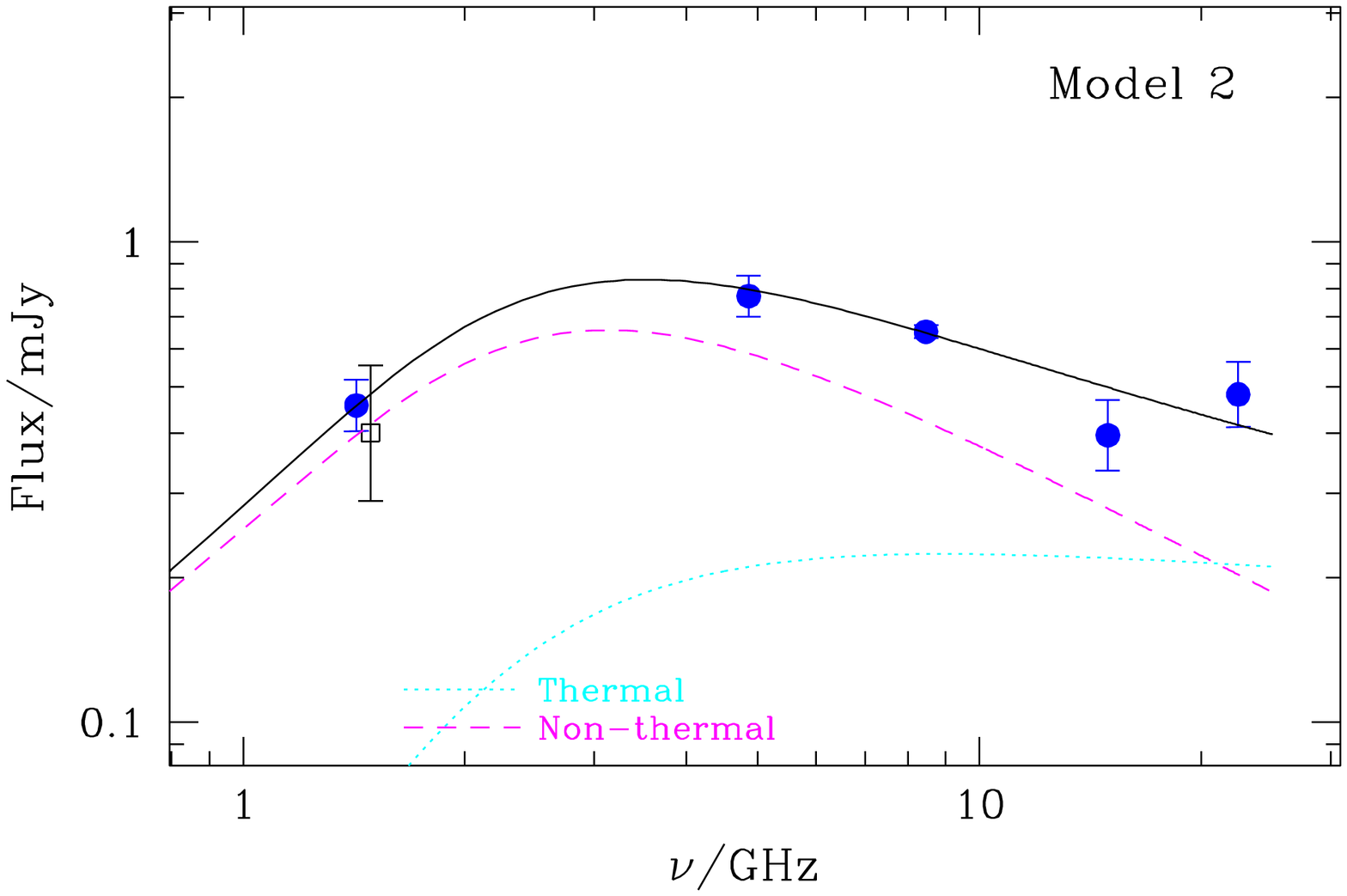}{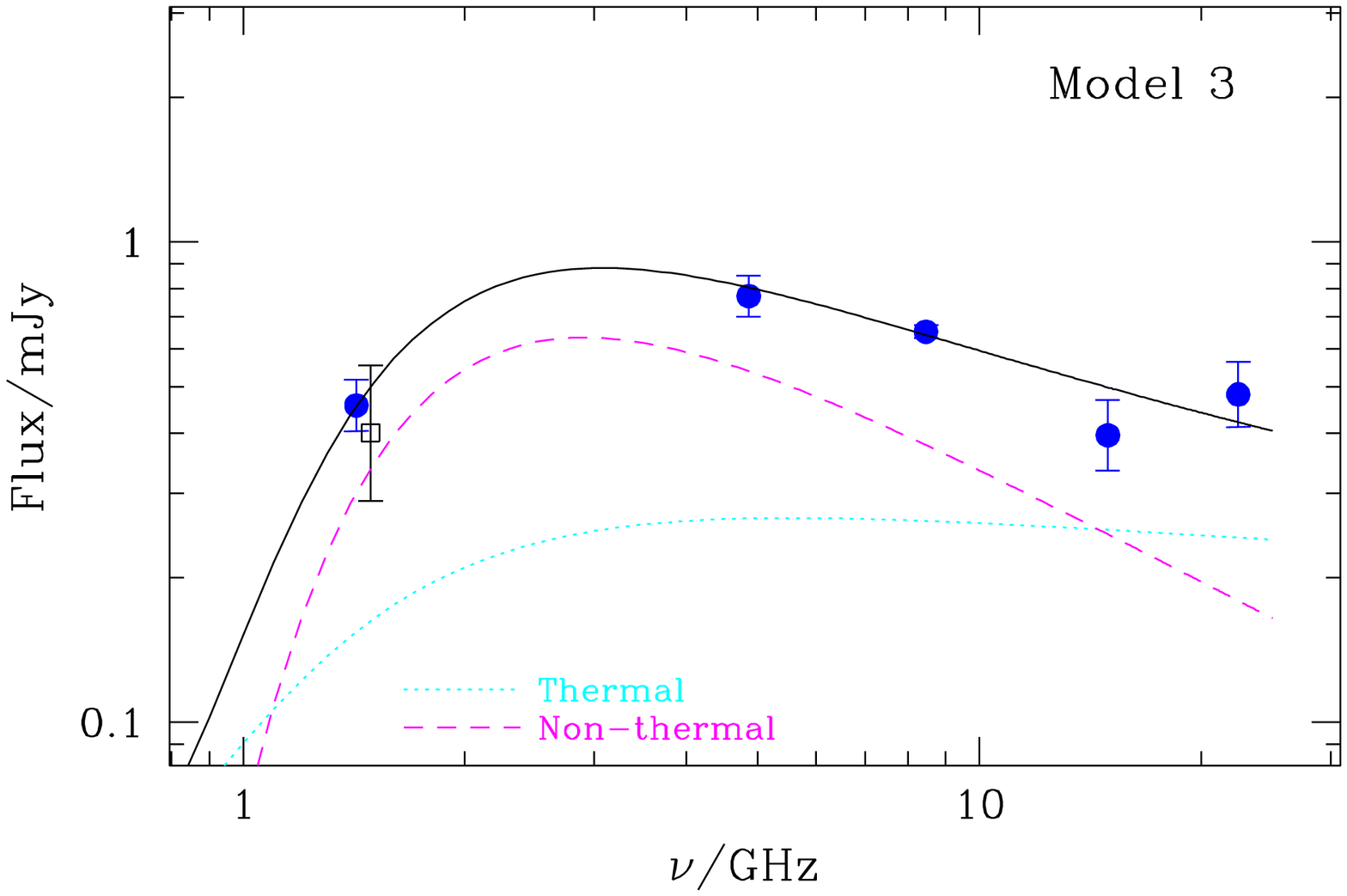}
}
\caption{Radio continuum flux versus frequency $\nu$.
Model 2 (mixed-geometry absorption of thermal$+$non-thermal components)
is shown in the left panel, and model 3 (mixed-geometry absorption
of thermal component $+$ screen absorption of non-thermal one) in the right.
The 1.49\,GHz point denoted by a filled square is from \citet{dale}.
A solid line shows the best-fit model, with the separate 
components given by a dotted line (thermal) and a dashed one (non-thermal). 
\label{fig:fit}}
\end{figure}

For $\alpha=-0.8$, the best fit for each of the three models 
gave a similar minimum of reduced $\chi_\nu^2=1.6$. 
Nominally, the best fit was obtained with the mixed geometry for both
components (model 2), but the lowest $\chi_\nu^2=1.60$ was only
slightly lower than that obtained with screen geometry for one or both
of the components (models 1 and 3: $\chi_\nu^2=1.67$).
The best-fit models together with the observed spectrum are shown
in Fig. \ref{fig:fit}.
With model 2 
(mixed geometry, left panel of Fig. \ref{fig:fit}), 
assuming a spherical geometry and
\alphant\,=\,$-0.8$, the best-fit parameters are:
\begin{eqnarray}
EM\,&=\,&(7.6\,\pm\,2.9) \times 10^7 \ {\rm pc\  cm^{-6}}\\
\theta\,&=\,&0.063\,\pm\,0.001 \  {\rm arcsec} \\
f^{\rm nt}_{\nuo}&=\,&0.68\,\pm\,0.007 \ {\rm mJy}
\end{eqnarray}
The resulting mean-square residuals are 0.06\,mJy, averaged over the 5 data points.
The thermal fraction at 5\,GHz is 0.27, which gives
a 5\,GHz thermal flux of 0.25\,mJy. 
With model 3 
(screen absorption for non-thermal, mixed for thermal,
right panel of Fig. \ref{fig:fit}) 
and a spherical geometry and
\alphant\,=\,$-0.8$, the best-fit parameters are:
\begin{eqnarray}
EM\,&=\,&(2.8\,\pm\,0.8) \times 10^7 \ {\rm pc\  cm^{-6}}\\
\theta\,&=\,&0.11\,\pm\,0.01 \  {\rm arcsec} \\
f^{\rm nt}_{\nuo}&=\,&0.60\,\pm\,0.005 \ {\rm mJy}
\end{eqnarray}
The thermal fraction at 5\,GHz is 0.32, which gives
a 5\,GHz thermal flux of 0.29\,mJy. 
Again, the resulting mean-square residuals are 0.06\,mJy.

There is an indication that the non-thermal spectral index is steeper
than the canonical $\alpha\,=\,-0.8$, since the lowest value of $\chi_\nu^2$ (1.37)
was obtained with model 2 by fixing $\alpha\,=\,-1.6$. 
However, while this result is formally a slightly better fit than that
we obtain with $\alpha\,=\,-0.8$, 
with such a steep $\alpha$ the other geometries give substantially worse fits.
Unlike steeper indices,
$\alpha\,=\,-0.8$ gives similarly low $\chi_\nu^2$ also for the different
absorption geometries.
We therefore chose to fix $\alpha\,=\,-0.8$. 

Our data are unable to distinguish among the various absorption
geometries (screen, mixed, or combined), although 
a screen absorption of both thermal and non-thermal components is 
physically unrealistic.
Lower frequency observations would be necessary to distinguish the 
models (see Fig. \ref{fig:fit}).
In what follows, we therefore consider the range of
the best-fit values implied by the mixed/combined geometries
(models 2 and 3).
The non-thermal fraction and flux levels do not vary much with
geometry (0.68$-$0.73, 0.60$-$0.68\,mJy, respectively), but the 
thermal emission measure and source size change by a factor
of two or more.

\subsection{Thermal Emission}

The fitted diameter of the thermal emission region $\theta$
corresponds to a physical size $D\,=\,17-29$\,pc,
several times 
smaller than the 80\,pc diameter from \citet{dale}
and the outer radius of 110\,pc predicted by the dust model of \citet{plante}.
Given $\theta$, the global $EM$ determined from our fit 
can be used to estimate the ionized gas density $n_e$. 
The resulting densities $n_e$ range from 980--2100\,cm$^{-3}$. 
These values are higher than those inferred from 
optical spectroscopy $n_e$ $\gtrsim$\,600\,cm$^{-3}$ \citep{izotov99}.
However in the optical we cannot probe very far into the star
clusters; hence the density could easily be higher than optical 
estimates.
Also, the optical spectroscopy is averaged over a larger region 
($\gtrsim$\,260\,pc) than the size we infer from our fit.

The best-fit emission measure of $2.8-7.6\times10^7$\,pc\,cm$^{-6}$
is not an extreme value for dwarf starburst galaxies.
In He\,2-10 and NGC\,5253, 
emission measures range from $10^7$ to $10^8$\,pc\,cm$^{-6}$
and implied densities between 500$\,<\,n_e\,<\,$10$^4$\,cm$^{-3}$.
\citep{kobulnicky,mohan}.
In Wolf-Rayet galaxies, typical emission measures
are $10^8-10^9$\,pc\,cm$^{-6}$, with
densities $\gtrsim\,10^3$\,cm$^{-3}$ \citep{beck2000}.
In II\,Zw\,40, another BCD,
\citet{beck2002} deduce an $EM$ of $10^9$\,pc\,cm$^{-6}$ 
for the most compact sources.
 
\subsection{Non-Thermal Emission}

Above 1.4\,GHz, the apparent spectral index is $-0.4$, a clear indicator of
significant non-thermal emission.
Indeed,
the fitted fraction of non-thermal radio emission at
$\nu\,=\,5$\,GHz is 0.7, with
a 5\,GHz non-thermal flux $f^{\rm nt}_{\nuo}\,=\,0.60-0.68$\,mJy. 
We can use $f^{\rm nt}_{\nuo}$ to estimate the magnetic field strength in \sbs,
assuming equipartition between the magnetic field energy and the
energy of the relativistic particles \citep{pach}.
We take the ratio of relativistic proton to electron energies to be
$k=40$ (e.g., \citealt{deeg}),
assume a spherical volume of diameter 17-29\,pc with unit filling factor,
and a constant spectral index from
$\nu_1\,=\,10^7$\,Hz to $\nu_2\,=\,10^{11}$\,Hz.
We obtain a synchrotron luminosity\footnote{$L_{\rm syn}$ is the 
total synchrotron luminosity obtained by integrating
the non-thermal spectrum from 
$\nu_1\,=\,10^7$\,Hz to $\nu_2\,=\,10^{11}$\,Hz.} 
\lsyn\,=\,$8-9\times10^{37}$\,erg\,s$^{-1}$, and
an equipartition $B\,=\,633-1065\,\mu$G, an
extremely high value for magnetic field strength.
The corresponding minimum pressure within the synchrotron emitting
region is $3\times10^{-8}-1\times10^{-7}$\,dyne\,cm$^{-2}$, and
the equipartition energy is $\sim10^{52}$\,erg.
These values, although extreme, are similar to 
the magnetic field strength and
pressure estimated for the compact sources in M\,82 \citep{allen}.

Although magnetic fields in very active star-forming galaxies
tend to be much stronger than in less active galaxies such as the Milky Way
\citep{kronberg},
we cannot be certain that 
such a high value for the magnetic field in \sbs\ is correct.
First, the source may be too young to have achieved the
minimum-energy configuration (see $\S$\ref{nonthermal}).
Second,
the standard formula for magnetic field strength
relies on a fixed integration interval in frequency.
In the case of \sbs, with its relatively high non-thermal radio 
luminosity, this procedure is almost certainly inappropriate.
A fixed frequency interval corresponds to different 
electron energy intervals, 
because of the dependence on magnetic field strength 
\citep{beck}.
Using the correct formula, and integrating from a proton
energy of 300\,MeV to infinity,
we obtain an equipartition field of 30\,$\mu$G (R. Beck, private
communcation), similar to the field strengths of other
BCDs \citep{deeg}.

\section{Radio Emission and Star Formation in \sbs\label{sfr}}

The results from the spectral fits
allow us to separate thermal and non-thermal emission in \sbs.
From the thermal emission,
we can then derive the number of ionizing photons, the star-formation rate,
and the supernova rate.

\subsection{Ionizing Photons \label{photons}}

With a thermal flux of $0.25-0.29$\,mJy at 5\,GHz, 
the inferred thermal luminosity
\lt\,=\,$8.8\times10^{19}-1.0\times10^{20}$\,W\,Hz$^{-1}$.
Because of the high electron temperature in \sbs, the usual
scaling for ionizing photons (e.g., \citealt{rubin,caplan,lequeux,condon}) 
give inconsistent results.
We therefore provide in the Appendix a new set of relations which are valid to within
0.5\% for 10000K$\leq T\leq$20000K.
These are useful for comparing thermal radio emission and optical/near-infrared
recombination lines in low-metallicity high temperature environments, and
we use them here.
With this thermal luminosity, Equation \ref{eqn:nion_radiolum} gives
the number of ionizing photons 
\nion\,=\,$8.2-9.4\times10^{52}$\,photons\,s$^{-1}$.
With 1 O7V star = 10$^{49}$\,photons\,\,s$^{-1}$ \citep{leitherer},
we obtain $\sim\,$9000 equivalent O7 stars.

From the \bra\ line luminosity of $3.2\times10^{39}$\,erg\,s$^{-2}$ \citep{hunt}
{\it uncorrected for extinction},
with $T=20000$K, we infer a total of 19400 stars. 
A correction for a screen extinction of
$A_V\simgt12$ \citep{hunt,plante} would increase these values by 
a factor of 1.6,
a mixed geometry correction a factor of 2.4.
It is clear that
the \bra\ is powered by between two and five times
the ionizing photons responsible for the thermal radio emission.

Either our estimate of the thermal fraction is wrong, or there is a source
of ionized gas emission in addition to the \hii\ region associated with
the SSCs.
The apparent spectral index $\alpha$ of $-0.4$ is indicative of a substantial
non-thermal component as mentioned before, and is also implied by our spectral fits.
Even if the entire observed 5\,GHz flux (0.8\,mJy) were thermal,  a 
reasonable extinction correction for the \bra\ emission would still give a 
similar discrepancy.
We therefore favor the hypothesis of an additional source of ionized gas,
namely an extended envelope of ionized gas resulting from massive stellar winds.
Such winds are known to exist in \sbs\ from its UV spectrum which shows
strong Si\,IV 1394, 1403 absorption lines with the characteristic P-Cygni
profiles \citep{thuan+izotov97}.

\subsection{The Stellar Wind \label{wind}}

To account for the presence of an extended stellar wind in \sbs,
we performed a new set of spectral fits which included
the spectrum of free-free radiation from varying-density envelopes
\citep{panagia},
in addition to the thermal and non-thermal components described above.
Traditional \hii\ region constant density models do not fit such
regions because of the different spectral behavior
($f_\nu \propto \nu^{0.6}$). 
Physically, such a configuration could occur if there were a 
density-bounded inner wind zone optically thick in the radio, 
but not in \bra, surrounded by an \hii\ region. 
The $\chi_\nu^2$ from these new fits
are formally equivalent to the best-fit models
of $\S$\ref{results}, although the influence of the wind in the radio
spectrum would not be apparent in Fig. \ref{fig:fit},
because of the low amplitude of the wind relative to the other components.
The best-fit model gives the fraction of the wind contribution at 5\,GHz as 
0.004, which corresponds to an 8.5\,GHz flux of 0.0035\,mJy.
With a cluster the size of that inferred in $\S$\ref{photons},
this would correspond to a stellar wind of the order of 
$\dot{M}_\odot\,\sim\,(v_\infty/1000 {\rm \,km\, s}^{-1})\, 
1\times10^{-4}$\,\msun\,yr$^{-1}$ per star (e.g., \citealt{lang}),
where $v_\infty$ is the stellar wind's terminal velocity.
This velocity being about 500\,km\,s$^{-1}$ in \sbs\
\citep{thuan+izotov97}, $\dot{M}_\odot$ is
roughly $5\times10^{-5}$\,\msun\,yr$^{-1}$.
This is not an unreasonable value compared to observed values for late-type
WN stars \citep{leitherer97} or for massive stars in the Galactic
center \citep{lang}.

We can predict how much \bra\ flux we would expect from
a stellar wind with the radio properties of \sbs.
Using the formalism in \citet{simon83},
and assuming that \bra\ is optically thick\footnote{If the
\bra\ flux were optically thin, the predicted flux would
be $10^5$ times higher than observed.
Also, these calculations assume an electron temperature
of 10000\,K.},
we would expect $3.2\times10^{-14}$ erg\,cm$^{-2}$\,s$^{-1}$,
several times higher than the observed \bra\ flux
of $9.0\times10^{-15}$ erg\,cm$^{-2}$\,s$^{-1}$ \citep{hunt}. 
An extinction correction for mixed geometry ($\tau_{\rm{Br}\alpha}=2.1$)
would give a flux of $2.3\times10^{-14}$ erg\,cm$^{-2}$\,s$^{-1}$,
lower than, but comparable to, the wind prediction.
The radio emission from the wind may be overestimated by our fit,
and the extinction correction for \bra\ uncertain,
but the wind$+$\hii-region model seems to be able to account
for the \bra\ flux better than the \hii-region model alone.

The radio contribution from the \hii\ region is far larger than that from the
stellar wind, while the \bra\ emission is much stronger from the wind
than from the \hii\ region. 
The \bra\ flux inferred from the thermal radio continuum,  
$3.7\times10^{-15}$ erg\,cm$^{-2}$\,s$^{-1}$, is
a factor of $\sim10$ times lower than what we would expect from an optically 
thick wind (see above). 
Moreover, it is less than half the observed value of 
$9.0\times10^{-15}$ erg\,cm$^{-2}$\,s$^{-1}$,
which is a lower limit because of the extinction correction. 
It is evident that in \sbs\ there is an excess of \bra\ flux over what 
we would expect from a simple \hii\ region.
The hot massive stars of the Arches Cluster
in the Galactic Center show a similar, if more extreme, excess;
comparing the radio emission \citep{lang} 
to the near-infrared recombination line flux \citep{nagata} shows
that the \bra\ luminosity is more than 20 times greater than expected
from a Case B \hii\ region extrapolated from the radio.

\subsection{Star-Formation Rate \label{subsfr}}

The thermal radio emission can also be used to
derive the massive star-formation rate (SFR).
First, using the formulation given in the Appendix, valid for high temperatures, 
and assuming that
$N({\rm He^+})/N({\rm H^+})\,=\,0.08$, $T\,=\,20000$K, and
$f_{{\rm H}\alpha}/f_{{\rm H}\beta}\,=\,2.75$,
from \lt\ we derive the \ha\ luminosity 
$L_{{\rm H}\alpha}\,=\,5.8-6.7\times10^{40}$\,erg\,s$^{-1}$.
Then, following \cite{condon}, we estimate the SFR of stars more massive
than 5\msun\ SFR$_{\geq5\msun}\,=\,0.13-0.15$\,\msun\,yr$^{-1}$.

The {\it total} SFR including less massive stars is $0.7-0.8$\,\,\msun\,yr$^{-1}$,
about half the value estimated by \citet{hunt}.
The reason for this is that the near-infrared recombination line flux is 
apparently partly due to stellar winds; consequently, 
numbers of massive stars and star-formation rates
inferred from the near-infrared recombination lines are 
overestimated.

We can infer the total stellar mass in the cluster, assuming an age
and a population synthesis model.
For an age of 5\,Myr \citep{papaderos,vanzi}, and with
Starburst 99 \citep{sb99}, the number of O7V
stars given by the thermal radio luminosity in \sbs\ corresponds to
a total stellar mass of 2.4$\times10^{6}$\msun.
This is very close to the total stellar mass 
of 2$\times10^{6}$\msun\ estimated by \citet{plante}
on the basis of the integrated infrared
spectral energy distribution, but is 3 times lower than that
of \citet{hunt}.
Again, this last discrepancy is because of the \bra\ excess described above.

\subsection{Non-thermal Radio Emission and Supernovae \label{nonthermal}}

Non-thermal radio emission can be used to estimate the star-formation
rate in \sbs, provided it follows usual scaling relations
(e.g., \citealt{condon}).
Here we investigate this assumption, and use our observations
to determine the origin of the non-thermal radio emission in \sbs.

Given the  SFR$_{\geq5\msun}$, we can estimate the supernova (SN) rate \snr.
Almost independently of the form of the initial mass function, 
\snr$\,=\,0.041\ {\rm SFR}_{\geq5\msun}$\,yr$^{-1}$.
With SFR$_{\geq5\msun}\,=\,0.13-0.15$\,\msun\,yr$^{-1}$, \snr\,=$0.005-0.006$\,yr$^{-1}$.
This is only 3--4 times smaller than the 
recent estimate of 0.016\,yr$^{-1}$ for M\,82 \citep{allen},
and the Type II SN rate of the Galaxy, 0.023\,yr$^{-1}$ \citep{tammann},
which however are galaxies over 1000 times more massive.
We can also use the relations given in \citet{condon}
for normal galaxies to deduce the non-thermal radio luminosity \lnt\ we would
expect from a typical starburst with this \snr.
In \sbs, predicted 
\lnt\ is $1.9-2.2\times10^{20}$\,W\,Hz$^{-1}$, corresponding to an expected
non-thermal flux $f_{\nuo}\,\approx\,0.6$\,mJy at 5\,GHz.
This value is very close to the model fit value of 0.60--0.68\,mJy at 5\,GHz.

It is puzzling why the scaling relation 
derived for massive spiral galaxies with evolved starbursts
(e.g., \citealt{condonyin}) would hold for \sbs.
Non-thermal radio emission in spiral galaxies is almost certainly produced
by  relativistic electrons diffusing over the galactic disk.
The timescale for this diffusion on kpc scales is $10^7-10^8$\,yr \citep{helou}.
As noted by \citet{condon}, although SNe almost certainly accelerate initially
the electrons responsible for non-thermal emission we observe,
$>$90\% of it was produced long after the discrete SNe have faded out.
The starburst in \sbs\ is very young, 5\,Myr or younger \citep{vanzi}.
It therefore seems unlikely that the diffusion mechanism active in the majority of
spiral galaxies has had enough time to operate
in \sbs\ to produce the observed amount of non-thermal emission.

The dimensions of the radio source as implied by the models
are also inconsistent with the standard diffusion picture.
The non-thermal source cannot be larger than the beam width (FWHM\,$\sim$400 pc),
and almost certainly is at least 10 times smaller than this given the 
constraints imposed by the spectral fit.
This size is 50--100 times smaller than that of the diffuse kpc-scale
radio emission mechanism which powers normal spiral galaxies.

We therefore investigate possible sources of non-thermal emission
which are both spatially compact and develop on a short timescale. 
On the very shortest timescales ($\simlt\,100$\,yr), 
the non-thermal radio emission in \sbs\ could come from
a single radio SN seen at a time not long after its explosion.
The inferred \lnt\,$=\,2\times10^{20}$\,W\,Hz$^{-1}$ is not unusual for radio SN
\citep{weiler},
but because of the rapid decline of radio emission
on timescales of decades, we must be observing it not long after it occurred.
However, the lack of radio variability over the 9-yr period spanned by our 
observations and those in \citet{dale} suggest that this interpretation may not
be the best one.

Alternatively, on slightly longer timescales,
the non-thermal radio emission in \sbs\ could arise either from
radio SNe or evolved SN remnants.
Radio SNe have adiabatic lifetimes of $\sim\,2\times10^4$\,yrs \citep{woltjer}, 
and from the \snr\ derived above, we can estimate the radio luminosity we would 
expect from a population of these (e.g., \citealt{ulvestad}).
From radio SNe
that emit only during the adiabatic phase of their evolution, we would expect
at 5\,GHz \lnt\,=\,3.5$\times10^{19}$\,W\,Hz$^{-1}$;
this is {\it 6 times smaller} than the
observed \lnt\,=\,$1.9-2.2\times10^{20}$\,W\,Hz$^{-1}$ at 5\,GHz.
The non-thermal source in \sbs\ is therefore difficult to interpret as 
an ensemble of standard radio SNe.

Finally, the radio source in \sbs\ could be interpreted as one or more evolved
compact SN remnants similar to those observed in M\,82 \citep{allen}.
As proposed by \cite{chevalier}, such remnants could be interacting with a dense
interstellar medium, with $n_e\,\sim\,10^3$\,cm$^{-3}$, similar to the $n_e$ we derive
for \sbs.
The timescales for this radiative phase are longer than the pure adiabatic lifetimes
\citep{chevalier},
and, given the intense starburst and massive SSCs in \sbs, we could expect to detect
sources with properties similar to those in M\,82.
Since \snr\ in \sbs\ corresponds to roughly 1 SN every 54\,yr,
it would not be unlikely for a few of these to be visible at the current epoch.
As mentioned before, the equipartition
pressure and magnetic field in \sbs\ are similar to those in M\,82.
The physical dimensions of such compact sources are $\simlt\,4$\,pc \citep{muxlow},
so they could conceivably reside
inside the 30\,pc diameter region responsible for the
thermal emission and absorption.
The non-thermal flux of \sbs\ placed at the 3.63\,Mpc distance of M\,82 would
be $\sim\,130$\,mJy, similar to the flux of the
most luminous of M\,82's compact sources, and 
equivalent to the flux of a few to ten or so of the less luminous ones.
\lsyn\ in \sbs\ is only 4 times higher than the most luminous source in M\,82,
and comparable to the predictions of \citet{chevalier}.
We conclude that a few compact evolved remnants, similar to those observed in 
M\,82 as well as other starbursts (e.g., \citealt{smith}), 
could be responsible for the non-thermal radio emission in \sbs.
VLBI and high-resolution X-ray observations are necessary to verify this hypothesis.

Indeed, Chandra observations of \sbs\ show evidence for 
a compact point source roughly spatially coincident with
the compact non-thermal radio source we find here,
and associated with the two brightest SSCs \citep{thuan-x}.
The X-ray source is extremely luminous in the 0.5$-$8.0\,keV range
($3\times10^{39}$\,erg\,s$^{-1}$),
and would be classified as an ultra-luminous X-ray source (ULX)
if it is a single object.

The nature of the radio$+$X-ray emitting object is however not
clear.
Whether the X-ray source is 
a single ULX or several high-mass X-ray binaries (HMXRBs)
is impossible to judge with the Chandra resolution.
But while there are more than a hundred HMXRBs in our Galaxy \citep{liu},
only $\leq$5\% of these are radio emitters \citep{clark}.
Although the X-ray luminosity is similar to that expected
from a microquasar associated with a medium-mass black
hole, its radio luminosity well exceeds (by $10^6$) that observed in
other such objects \citep{mirabel}.
Also its non-thermal radio spectrum is much steeper than the flat
spectra thought to be characteristic of black hole systems in the
low/hard spectral state \citep{marti}.
Moreover, as mentioned above, there 
is no evidence for either radio variability over a 9-yr interval or
X-ray variability over shorter timescales,
although the limited photon statistics do not strongly constrain 
the temporal nature of the X-ray source \citep{thuan-x}.
Hence, the radio source and the X-ray one could be physically distinct
objects, 
the non-thermal radio luminosity coming from evolved
SNRs expanding in a dense medium, 
and the X-ray luminosity from a single ULX or several HMXRBs.
Independently of the specific nature of the compact object(s) 
responsible for the X-ray and radio emission, it is evident that they 
are connected to the SSCs and the starburst episode which formed them. 

\subsection{Implications for High-Redshift Star Formation}

There are two noteworthy features of the radio continuum in \sbs,
relative to spectra of normal spiral galaxies or other BCDs.
First, there is significant free-free absorption on a global scale,
of sufficient amplitude to suppress by a factor of 4 or more the 
observed flux at 1.49\,GHz.
Second, while the spectrum of \sbs\ is significantly different from
those of normal spirals, it is not completely thermal as in other BCDs.
At 5\,GHz, we find roughly a 30-70 mix of thermal and non-thermal
emission.

If \sbs\ can be taken as representative of star formation in an
extremely low-metallicity environment,
care should be taken when deriving cosmological conclusions
from present and future radio continuum surveys.
Because of the young age, low metal abundance, and intensity of
the starburst,
galaxies like \sbs\ may not obey the usual scaling laws
laid out in \citet{condon}.
Moreover, the effects of free-free absorption substantially alter
the spectrum at frequencies $\simlt\,$1.5\,GHz.
It would not be straightforward to disentangle such an effect
when deriving the SFR as a function of redshift. 
Young starbursts occurring in a metal-poor, but dusty and dense
environment, have integrated radio properties that may be
very different from  those of evolved starburst prototypes 
such as M\,82 and Arp\,220.
As a result, photometric redshifts inferred from ``standard''
spectral-energy distributions could also be incorrect.

\section{Summary and Conclusions}

We have presented new VLA continuum observations of the extremely
metal-poor BCD \sbs, and fit the spectrum to a variety of models.
\begin{itemize}
\item
The best-fitting model (model 2),
which assumes an absorption medium intermixed with thermal$+$non-thermal
emission, gives an $EM=7.6\times10^7$\,pc\,cm$^{-6}$
and a diameter of the radio-emitting region of 17\,pc. 
The inferred density is $n_e\,\sim\,2000$\,cm$^{-3}$.
\item
The non-thermal fraction at 5\,GHz is 0.7, with an unabsorbed
non-thermal 5\,GHz flux of 0.68\,mJy, corresponding to
\lnt\,=\,2.2$\times10^{20}$\,W\,Hz$^{-1}$.
From this, we derive an equipartition magnetic field of $\sim$\,1\,mG,
and a pressure of $\sim10^{-7}$\,dyne\,cm$^{-2}$.
Integrating over a fixed electron energy interval, rather than one
at fixed frequency, gives a field strength of 30$\,\mu$G.
\item
We find evidence for a stellar wind in \sbs, because of the excess
of \bra\ flux over that inferred from the thermal radio emission.
\item
Because of the compact size and young age of the starburst,
it is difficult to interpret the
non-thermal radio emission as due to 
diffusion of SN-accelerated electrons on timescales of $10^7-10^8$\,yr.
Rather, 
we attribute the non-thermal radio emission to an
ensemble of compact SN remnants expanding in a dense interstellar medium.
\end{itemize}

\acknowledgments

KKD is supported by a National Science Foundation Astronomy
and Astrophysics Postdoctoral Fellowship under award AST-0103879.
TXT acknowledges the partial
financial support of NSF grant AST-0205785.
LKH would like to thank Rainer Beck,
Riccardo Cesaroni, Dick Crutcher, and Marcello Felli for
extremely enlightening discussions.
This paper has made use of the astrometry package WCS Tools available from
http://tdc-www.harvard.edu/software/wcstools.html.

\appendix

\section{Ionizing Photons and Optical/Radio Ratio for Electron Temperatures $>$ 10000\,K}

For a radiation-bounded \hii\ region, 
\begin{equation}
N_{\rm ion}\,=\,\int n_e n(H^+) \alpha_B dV 
\label{eqn:nion}
\end{equation}
where $N_{\rm ion}$ is the number of ionizing photons,
$n_e$ and $n(H^+)$ are the electron and ionized hydrogen 
number densities
respectively,
and $\alpha_B$ is the Case B recombination coefficient
\citep{osterbrock}.
Here, we have neglected the absorptions by He since they do not
greatly reduce the number of photons available for ionizing H 
\citep{osterbrock}.

The radio free-free emissivity $j_\nu$ can be written as:
\begin{equation}
4 \pi j_\nu\,=\,4.102\times10^{-39} \left(\frac{\nu}{GHz}\right)^{-0.1}
\left(\frac{T}{10^4\,K}\right)^{-0.35} n_e [n({\rm H}^+) + n({\rm He}^+)]
\label{eqn:radio_emissivity}
\end{equation}
in units of erg s$^{-1}$ cm$^{-3}$ Hz$^{-1}$, and
where $T$ is the electron temperature,
and $n({\rm He}^+)$ is the number density of ionized helium 
\citep{altenhoff,mezger,rubin,caplan}.
The correction factor between this approximation and the more exact form given
in \cite{oster} is 0.9975 for $T=20000$\,K (see \citealt{mezger}).
Integrating $4 \pi j_\nu$ over the emitting volume and placing a source
with this luminosity at distance $d$ gives an expression for the
radio flux:
\begin{equation}
f_\nu\,=\,3.43\times10^{-63} \left(\frac{d}{Mpc}\right)^{-2}
\left(\frac{\nu}{GHz}\right)^{-0.1} \left(\frac{T}{10^4\,K}\right)^{-0.35} 
\int n_e [n({\rm H}^+) + n({\rm He}^+)] dV \quad mJy
\label{eqn:flux}
\end{equation}

The temperature dependence of the recombination coefficient $\alpha_B$ 
is very different for temperatures $>$ 10000\,K, than for cooler ones.
If a power law approximation is assumed for this dependence,
and using the values given by \citet{osterbrock}, 
for $T$ 10000\,K$\leq T\leq$20000\,K we find:
\begin{equation}
\alpha_B \,=\, 2.59\times10^{-13} \left( \frac{T}{10^4\,K} \right)^{-0.04}   .
\label{eqn:alpha}
\end{equation}

By combining Equations \ref{eqn:flux} and \ref{eqn:nion},
and eliminating the volume integrals, we can 
write the number of ionizing photons $N_{\rm ion}$ as a function of
thermal free-free flux:
\begin{equation}
N_{\rm ion}\,=\,7.56\times10^{49} \left(\frac{d}{Mpc}\right)^{2}
\frac{n({\rm H}^+)}{n({\rm H}^+) + n({\rm He}^+)} 
\left(\frac{\nu}{GHz}\right)^{0.1} \left(\frac{T}{10^4\,K}\right)^{0.31} 
\left(\frac{f_\nu}{mJy}\right)
\label{eqn:nion_radio}
\end{equation}
Except for the different temperature dependence,
this equation is the same as \citet{rubin} and \citet{lequeux}, 
but we have left explicit the effect of ionized He; 
$n({\rm He}^+)/n({\rm H}^+)$ is $\sim 0.08$ for low-metallicity
environments (e.g., the Magellanic Clouds: \citealt{caplan};
 \sbs: \citealt{melnick}). 

Equation \ref{eqn:nion_radio} can also be expressed in terms of radio
luminosity $L_T$ rather than flux:
\begin{equation}
N_{\rm ion}\,=\,6.32\times10^{52} 
\frac{n({\rm H}^+)}{n({\rm H}^+) + n({\rm He}^+)} 
\left(\frac{\nu}{GHz}\right)^{0.1} \left(\frac{T}{10^4\,K}\right)^{0.31} 
\left(\frac{L_T}{10^{20} W Hz^{-1}}\right)
\label{eqn:nion_radiolum}
\end{equation}
Equation \ref{eqn:nion_radiolum} is equivalent to that given by 
\citet{condon}, except for the different temperature dependence
and the explicit inclusion of the dependence on He$^+$;
our expression is valid for 10000\,K$\leq T \leq$20000\,K.

We now turn to the hydrogen recombination lines for 10000\,K$\leq T \leq$20000\,K.
Using the coefficients in \citet{osterbrock}, and assuming a power-law dependence
on $T$, the (Case B) \hb\ emissivity $j_{H\beta}$ can be written as:
\begin{equation}
4\pi j_{H\beta} = 1.24\times 10^{-25} n_e n(H^+) \left(\frac{T}{10^4\,K}\right)^{-0.91} 
\label{eqn:hbeta_emissivity}
\end{equation}
in units of erg s$^{-1}$ cm$^{-3}$ (c.f., \citealt{mezger}).
As before, we integrate $4\pi j_{H\beta}$ over the emitting volume to obtain
the total $H\beta$ luminosity $L_{H\beta}$, which, when combined with 
Eq. \ref{eqn:nion} and Eq. \ref{eqn:alpha}, gives the number of ionizing photons
as a function of $L_{H\beta}$:
\begin{equation}
N_{\rm ion}\,=\,2.09\times10^{12} \left(\frac{T}{10^4\,K}\right)^{0.87} 
\left(\frac{L_{H\beta}}{erg\,s^{-1}}\right)
\label{eqn:nion_hbeta}
\end{equation}
Placing this source at distance $d$ gives an expression which depends on the
\hb\ flux $f_{H\beta}$:
\begin{equation}
N_{\rm ion}\,=\,2.50\times10^{50} \left(\frac{T}{10^4\,K}\right)^{0.87} 
\left(\frac{d}{Mpc}\right)^{2}
\left(\frac{f_{H\beta}}{10^{-12}\, erg\,cm^{-2}\,s^{-1}}\right)
\label{eqn:nion_hbetaflux}
\end{equation}

The \hb\ flux expected from the radio free-free emission is therefore given by:
\begin{equation}
\left(\frac{f_{H\beta}}{10^{-12}\, erg\,cm^{-2}\,s^{-1}}\right)\,=\,
0.302\ 
\frac{n({\rm H}^+)}{n({\rm H}^+) + n({\rm He}^+)} 
\left(\frac{T}{10^4\,K}\right)^{-0.56} 
\left(\frac{\nu}{GHz}\right)^{0.1}
\left(\frac{f_\nu}{mJy}\right)
\label{eqn:radio_hb}
\end{equation}
Equation \ref{eqn:radio_hb} is very similar to \citet{condon} because the
different temperature dependence on the recombination coefficient used here 
disappears in the ratio.

Other recombination lines can be used if the temperature dependence is
taken into account.
Again, using the coefficients in \citet{osterbrock}, we find for
10000\,K$\leq T \leq$20000\,K:
\begin{eqnarray}
j_{H\alpha}/j_{H\beta}   & = & 2.86\ \left(\frac{T}{10^4\,K}\right)^{-0.057} \\
j_{H\beta}/j_{Pa\alpha}  & = & 2.959\ \left(\frac{T}{10^4\,K}\right)^{0.251} \\
j_{H\beta}/j_{Br\gamma}  & = & 35.939\ \left(\frac{T}{10^4\,K}\right)^{0.2454} \\
j_{H\beta}/j_{Br\alpha}  & = & 12.496\ \left(\frac{T}{10^4\,K}\right)^{0.352}
\end{eqnarray}
These expressions are valid for densities 100\,cm$^{-3}\leq n_e \leq$1000\,cm$^{-3}$.

\clearpage

\end{document}